\documentclass{osa-article}

\journal{osajournal}

\articletype{Research Article}
\usepackage{amsmath,geometry,times,float,graphicx,color,textcomp}
\begin{document}

\title{Circumventing the optical diffraction limit with customized speckles}
\author{Nicholas Bender,\authormark{1} Mengyuan Sun,\authormark{2} Hasan Y{\i}lmaz,\authormark{1} Joerg Bewersdorf,\authormark{3,4} and Hui Cao\authormark{1,*}}
\address{\authormark{1} Department of Applied Physics, Yale University, New Haven CT 06520, USA\\
\authormark{2} Department of Molecular Biophysics and Biochemistry, Yale University, New Haven CT 06520, USA \\
\authormark{3} Department of Cell Biology, Yale University School of Medicine, New Haven CT 06520, USA\\
\authormark{4} Department of Biomedical Engineering, Yale University, Yale University, New Haven CT 06520, USA
}

\email{\authormark{*} hui.cao@yale.edu} 

\begin{abstract}
Speckle patterns have been widely used in imaging techniques such as ghost imaging, dynamic speckle illumination microscopy, structured illumination microscopy, and photoacoustic fluctuation imaging. Recent advances in the ability to control the statistical properties of speckles has enabled the customization of speckle patterns for specific imaging applications. In this work, we design and create special speckle patterns for parallelized nonlinear pattern-illumination microscopy based on  fluorescence photoswitching. We present a proof-of-principle experimental demonstration where we obtain a spatial resolution three times higher than the diffraction limit of the illumination optics in our setup. Furthermore, we show that tailored speckles vastly outperform standard speckles. Our work establishes that customized speckles are a potent tool in parallelized super-resolution microscopy.
\end{abstract}
\section*{Introduction}

When a large number of random partial waves interfere, an irregular pattern of speckled grains, commonly called a speckle pattern or speckles, is created.  If the amplitudes and phases of the partial waves independently vary -and the phases are uniformly distributed over a range of $2 \pi$- the speckle pattern is fully developed \cite{DaintyB, GoodmanB, freund1001}. Most speckle patterns satisfy Rayleigh statistics, wherein the joint probability distribution of the complex field is circular Gaussian, and the probability density function (PDF) of the intensity is a negative exponential \cite{GoodmanB, DaintyB}. The topology of a Rayleigh speckle pattern can be characterized as an irregular and interconnected low-intensity web with isolated high-intensity islands. While the bright speckle grains are all relatively isometric and isotropic \cite{freund1001}, the dark regions surrounding optical vortices are anisotropic \cite{berry2000phase, dennis2001, WangPRL2005, GenackPRL2007}. In various imaging and metrology techniques speckle patterns are used to illuminate samples in order to improve precision, resolution, visibility, and sectioning while simultaneously providing a wide field of view \cite{takedaSPIE03, shapiro2008computational, katz2009compressive, bromberg2009ghost, Dynamic2,  mertz2011optical, Dynamic4, 2012_Sentenac, 2013_Min_SciRep, oh2013sub, 2015_Kim_SciRep, 2014_Zheng_super_resolution, yilmaz2015speckle,   2017_Waller_super_resolution}.

In fluorescence microscopy, a sub-diffraction-limited resolution can be obtained with saturable structured illumination techniques. Examples include stimulated emission depletion (STED) microscopy \cite{hell1994breaking, hell2007far}, ground state depletion (GSD) microscopy \cite{ hell1995ground}, and reversible saturable optical fluorescence transitions (RESOLFT) \cite{grotjohann2011diffraction, hofmann2005breaking, chmyrov2013nanoscopy}. Generally, these techniques rely on spatially modulating the illumination intensity to toggle fluorescence on and off in a spatially selective manner. For example, in RESOLFT, a doughnut-shaped optical beam photoconverts all fluorophores in a region of a sample except those close to the vortex center, and therefore fluorescence is only emitted from a sub-diffraction-sized region. Usually, point scanning of the doughnut beam is required to construct an image which can be very time consuming. Recently, this process has been parallelized using either a one-dimensional (1D) standing wave pattern \cite{Rego012PNAS} or a two-dimensional (2D) lattice of doughnuts \cite{chmyrov2013nanoscopy}. The first method involves rotating or translating the illumination pattern, and as a result, post-processing is required for 2D super-resolution since each individual pattern will only improve the resolution in one direction. In the second method, two incoherently-superimposed orthogonal standing waves create a square lattice of intensity minima, but axial uniformity of the intensity pattern hinders optical sectioning. One way to circumvent this issue is to use the optical vortices in a speckle pattern as the nonlinear structured illumination pattern \cite{2016_Guillon_PRL}. Because fully-developed speckle patterns rapidly and non-repeatably change along the axial direction, they can be used to obtain a three-dimensional super-resolution image \cite{2019_Guillon_NatCommun}. In this context, speckles are advantageous because they are robust to optical distortions/aberrations while simultaneously enabling three-dimensional super-resolution \cite{shapiro2008computational, katz2009compressive, bromberg2009ghost, Dynamic2,  mertz2011optical, Dynamic4, 2012_Sentenac, 2013_Min_SciRep, oh2013sub, 2015_Kim_SciRep, 2014_Zheng_super_resolution, yilmaz2015speckle,   2017_Waller_super_resolution, 2019_Guillon_NatCommun,meitav2016point}. The problem with using Rayleigh speckles, however, is that the anisotropic and strongly fluctuating shapes of their optical vortices leads to non-isotropic and non-uniform improvements in the spatial resolution.

In this work, we introduce and use an ideal family of tailored speckle-patterns for nonlinear pattern-illumination microscopy. The speckle patterns we design need to possess the following properties in order to produce isotropic and uniform super-resolution. All of the vortices in the speckles must have a circular shape and an identical high-intensity halo surrounding each vortex core. Away from the randomly distributed optical vortices, however, the spatial intensity profile of the speckles should be uniform with minor fluctuations. When this is the case, the speckles’ intensity PDF is a narrow peak that can be approximated as a delta function. We thus refer to this family of tailored speckles as "delta" speckles. To enable optical sectioning, the delta speckles' field must be fully developed and as a result the speckles’ intensity pattern will rapidly and non-repeatably evolve upon axial propagation. Experimentally, we generate delta speckle patterns by modulating the phase-front of a monochromatic laser beam with a spatial light modulator. Using delta speckle patterns, we photoconvert a planar fluorescent gel sample and measure the fluorescence signal from the unconverted regions in the vicinity of optical vortices. Not only do we demonstrate a $3\times$ enhanced spatial resolution relative to the optical diffraction limit of the illumination optics, but also, we show the significant advantages of using customized speckles over Rayleigh speckles in fluorescence microscopy.

\begin{figure*}[ht]
\begin{center}
\includegraphics[width=\linewidth]{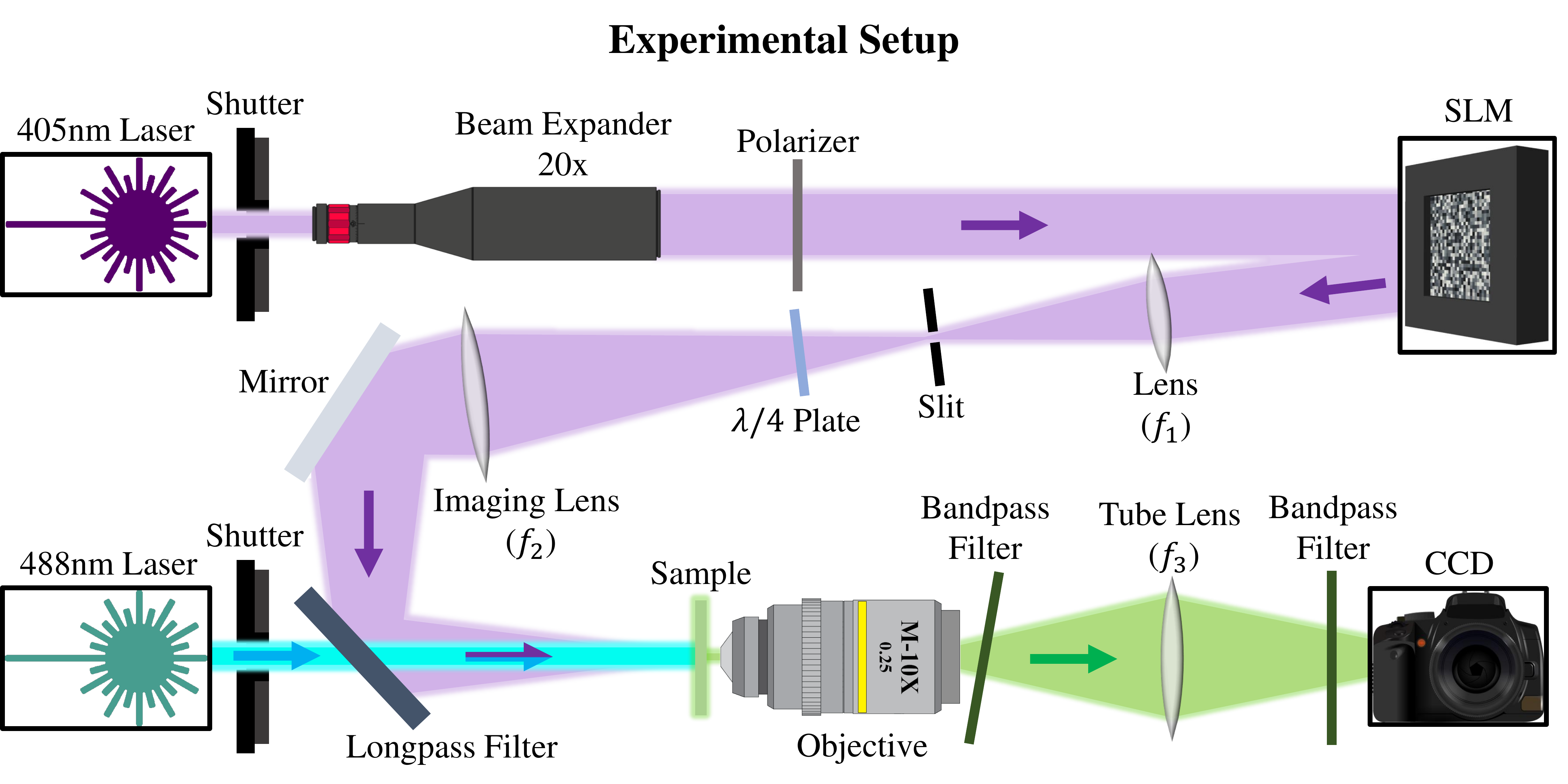}
\caption{\label{Setup} {\bf Experimental setup.} The experimental setup used for nonlinear speckle-illumination microscopy is shown. We use a phase-only SLM to generate a customized speckle pattern which illuminates and photoconverts a fluorescent sample.} 
\end{center}
\end{figure*}

\section*{Experimental Setup}
In our demonstration, the fluorescent sample is a 25-\textmu m thick layer of the photoconvertible protein mEos3.2, which is uniformly distributed and suspended in a gel \cite{zhang2012rational} (see the Supplementary Materials for details). A continuous-wave (CW) laser beam with a wavelength of $\lambda=405$ nm is used to photoconvert the protein. The laser’s wavefront is modulated by a phase-only spatial light modulator (SLM), which generates a speckle pattern in its far-field (Fourier plane). The polarization of the far-field speckles’ field is converted from linear to circular by a quarter-waveplate. The far-field speckle pattern illuminates and photoconverts the sample.

A schematic of the detailed experimental setup for photoconverting a fluorescent sample with customized speckle patterns and imaging the unconverted fluorescence is shown in Fig.~\ref{Setup}. The $\lambda=405$ nm CW laser beam is expanded and linearly polarized before it is incident on the phase-only SLM (Meadowlark Optics). The pixels on the SLM can modulate the phase of the incident field between $0$ and $2 \pi$: in increments of $2 \pi / 90$. Because a small portion of light reflected from the SLM is unmodulated, we write a binary phase diffraction grating on the SLM and use the light diffracted to the first order for photoconversion. In order to avoid cross-talk between the neighboring SLM pixels, $32 \times 32$ pixels are grouped to form one macropixel, and the binary diffraction grating is written within each macropixel with a period of 8 pixels. We use a square array of $32 \times 32$ macropixels in the central part of the phase modulating region of the SLM to shape the photoconverting laser light. The light modulated by our SLM is Fourier transformed by a lens with a focal length of $f_{1}=500$ mm and cropped in the Fourier plane with a slit to keep only the first-order diffraction. The complex field on the Fourier plane of the SLM is imaged onto the surface of the sample by a second lens, with a focal length of $f_{2}=500$ mm. Using a $\lambda/4$ plate, we convert the linearly polarized photoconverting beam into a circularly polarized beam before it is incident upon the sample. In this setup, the full width at half-maximum of a diffraction-limited focal spot is 17 \textmu m.

After the photoconversion process, a second laser, operating at a wavelength of $\lambda = 488$ nm, uniformly illuminates the sample and excites the non-photoconverted mEos3.2 proteins. The unconverted protein fluorescence has a wavelength centered around $\lambda=532$ nm \cite{zhang2012rational}.
To collect the fluorescence, we use a $10\times$ objective of ${\rm NA}=0.25$ and a tube lens with a focal length of $f_{3}=150$ mm. The 2D fluorescence image is recorded by a CCD camera (Allied Vision Manta G-235B). The spatial resolution of the detection system is estimated to be $1.1$ \textmu m, which allows us to resolve features in the sample that are smaller than the diffraction limit of our illumination optics. The remaining excitation laser light, which is not absorbed by the sample, is subsequently removed by two Chroma ET 535/70 bandpass filters. One filter is placed after the objective lens and reflects the excitation beam off the optical axis of our system. The second one is placed directly in front of the camera.

\section*{Photoconversion Fluorescence Microscopy}
\begin{figure*}[t]
\centering
\includegraphics[width=\linewidth]{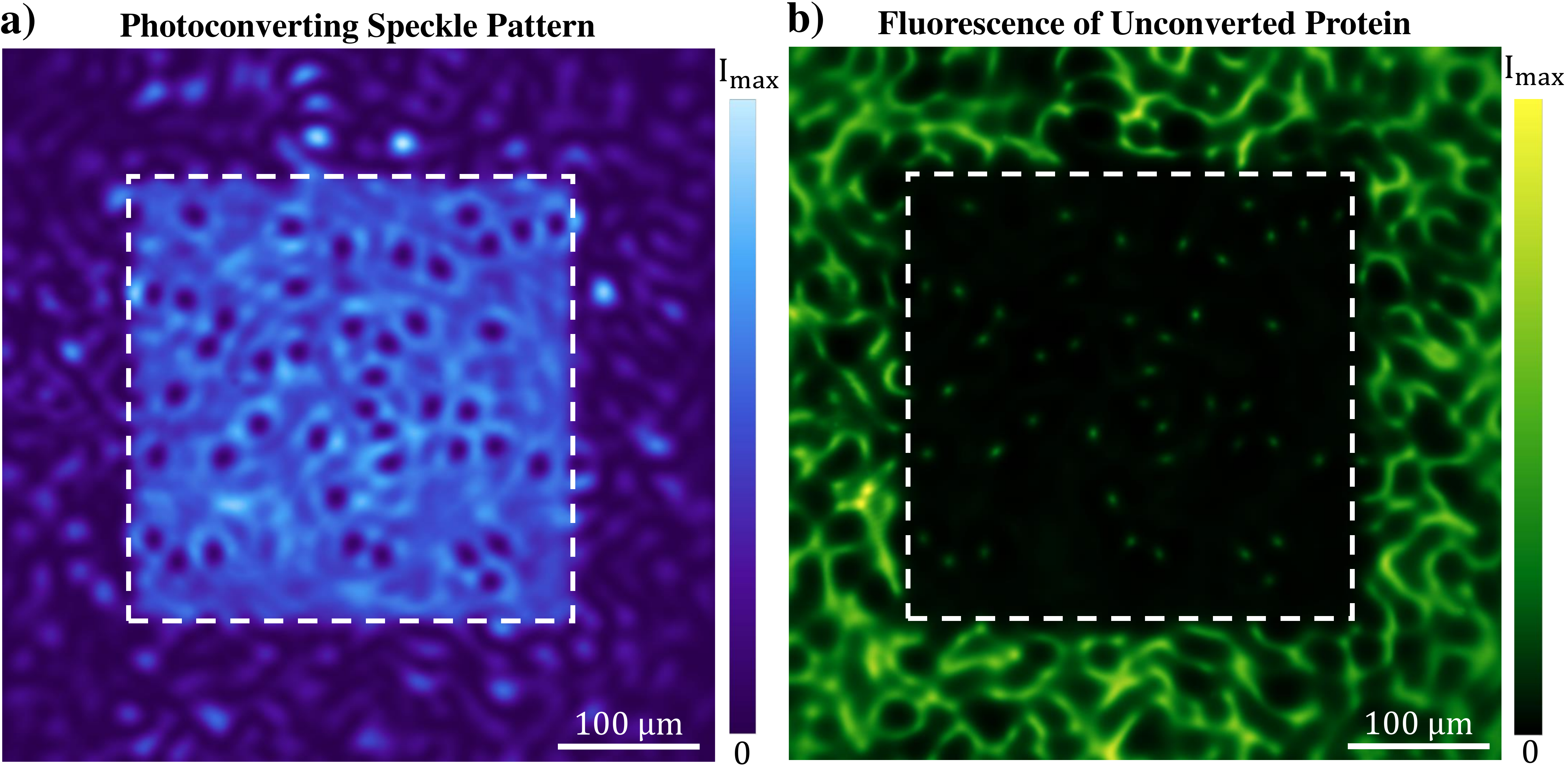}
\caption{\label{Demo} {\bf Customizing speckles for nonlinear pattern illumination.} An experimentally recorded speckle pattern that illuminates and photoconverts a uniform fluorescent protein sample is shown in (a). Within the white box the speckles obey delta statistics and outside they obey Rayleigh statistics. In (b), an experimentally recorded image of the fluorescence from the unconverted regions shows isometric and isotropic spots produced by the vortices in the delta speckles; while the region photoconverted by the Rayleigh speckles features large, irregular, and interconnected fluorescent grains.} 
\end{figure*}
Fig.~\ref{Demo}(a) shows part of an example speckle pattern used to photoconvert the fluorescent sample. The lateral dimension of the entire speckle pattern is 600 \textmu m. The speckle pattern consists of two regions. Within the central region marked by the white dashed square, 300 \textmu m $\times$ 300 \textmu m in size, the speckle pattern obeys delta intensity statistics, i.e. a random array of circular vortices embedded in a nearly uniform intensity background. Outside the square, the speckles adhere to Rayleigh statistics, featuring sparse, nearly circular islands of high intensity, randomly distributed in a dark sea. The average speckle grain size is 17 \textmu m. The speckles are created using a modification of the methods presented in Refs. \cite{benderAPL2019, benderOE2019, bender2018} (see the Supplementary Materials for details). 

\begin{figure*}[ht]
\centering
\includegraphics[width=\linewidth]{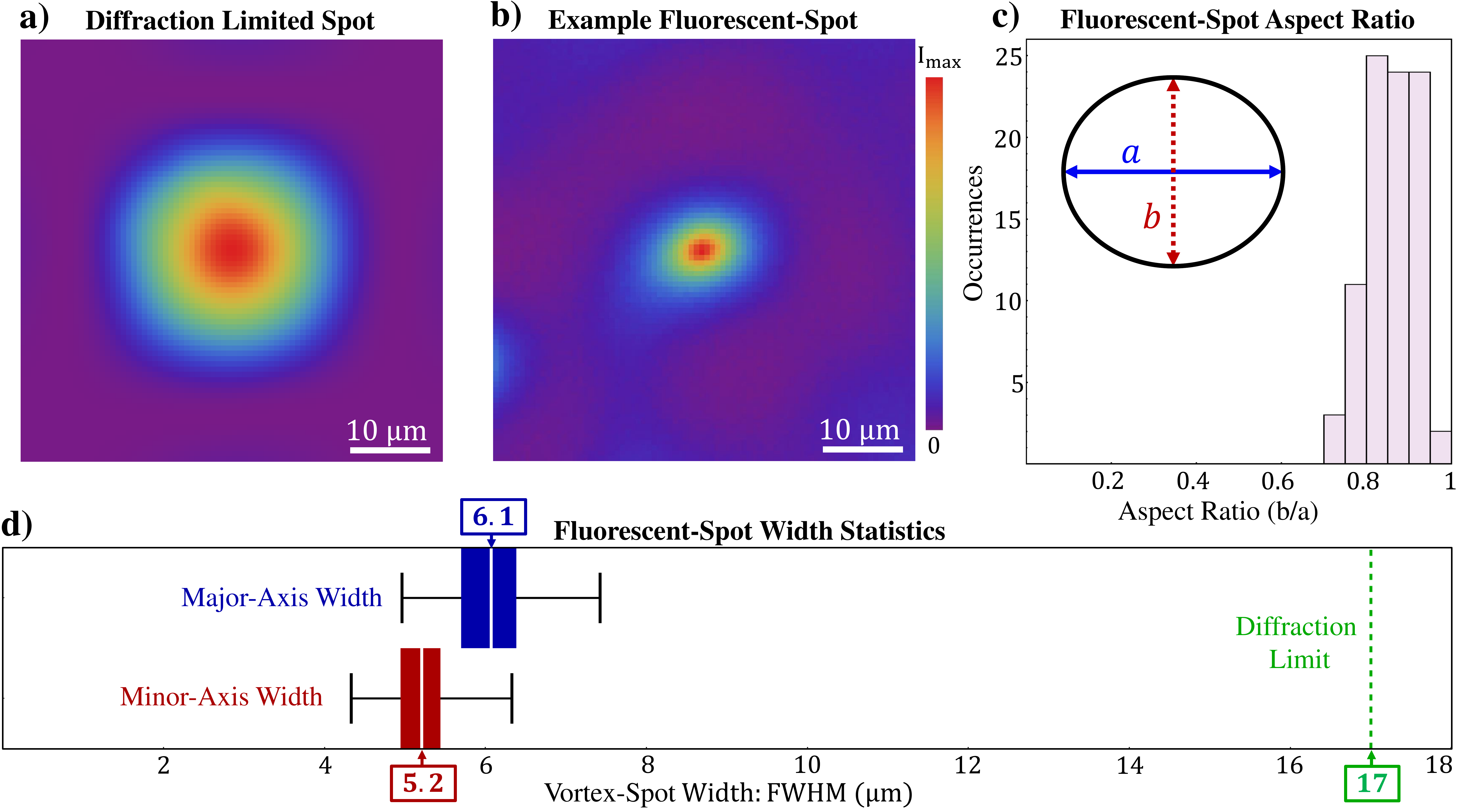}
\caption{\label{PSF} {\bf Circumventing the diffraction limit.} A diffraction-limited spot of the 405 nm illumination/photoconversion optics is presented in (a). In (b), an example fluorescent spot produced by a vortex in the delta speckle pattern is illustrated. Its size is much smaller than the diffraction-limited spot. Its shape can be fit by an ellipse of major axis width $a=6.1$ \textmu m and minor axis width $b=5$ \textmu m. The aspect-ratio, $b/a$, histogram of all fluorescent spots produced by the delta speckles is shown in (c). The inset illustrates an ellipse with the average aspect ratio $\langle b/a \rangle =0.86$. In (d), the box-plot analysis of the major and minor axes widths is shown. The white line marks the mean value, and the black whiskers represent the upper and lower bounds of the data. The edges of the blue and red shaded regions mark the upper and lower quartiles ($25\%,75\%$) of the ensemble. The green dashed line indicates the FWHM of the diffraction-limited spot in (a).}
\end{figure*}

 Fig.~\ref{Demo}(b) shows the fluorescence from the uniform protein gel sample after being photoconverted by the speckle pattern presented in Fig.~\ref{Demo}(a). Outside the central square, the sample is photoconverted by the Rayleigh speckle pattern. The fluorescent pattern consists of a sprawling anisotropic web, which reflects the topology of the low-intensity regions surrounding the optical vortices. In stark contrast, the fluorescence pattern within the central square, which is photoconverted by delta speckles, features isolated fluorescent spots of a much smaller size. They are all created by the optical vortices in the central square of Fig.~\ref{Demo}(a). The high level of isotropy exhibited by the fluorescent spots originates from the high degree of rotational symmetry that is present in the vortices generating them. Apart from these spots, the fluorescent intensity is uniformly low. This dark background arises from the homogeneity of the delta speckle pattern’s intensity away from optical vortices. A qualitative comparison between the distinct fluorescence patterns inside and outside the central square illustrates how customizing the intensity statistics of speckles can significantly enhance the performance of speckle-based pattern illumination microscopy.    

Next, in Fig.~\ref{PSF}, we quantitatively assess the performance of delta speckles in our application. We compare a diffraction-limited spot, shown in Fig.~\ref{PSF}(a), with the fluorescent spots produced in the central region of Fig.~\ref{Demo}(b) to quantitatively determine the resolution enhancement. The diffraction-limited spot is created using the experimentally measured field-transmission matrix of the 405 nm light, from the SLM to the sample, and it gives the 2D point spread function (PSF) of our illumination optics. The full width at half maximum (FWHM) of the spot is 17 \textmu m, which is in agreement with the estimated value of 18 \textmu m from the diffraction limit $\lambda/(2 {\rm NA})$,  where ${\rm NA} \cong 0.011$ is the numerical aperture of the illumination optics. Fig.~\ref{PSF}(b) shows an example fluorescent spot created by one of the vortices in the illuminating delta speckle pattern. Its shape is close to circular. We determine the full width at half maximum (FWHM) along both the major axis $a$ = 6.1 \textmu m and the minor axis $b$ = 5.0 \textmu m by fitting (see the Supplementary Materials for details). The average spot size, $(a+b)/2 \cong$ 5.6 \textmu m, is three times below the diffraction limit (17 \textmu m).  

To determine the isotropy of the spatial resolution enhancement produced by the delta speckles, we measure the major axis $a$ and minor axis $b$ of all fluorescent spots in the central region.  The aspect ratio $b/a$ is a measure of isotropy. Using measurements from a total of 89 fluorescent spots stemming from two independent delta speckle patterns, we calculate a histogram of the ratio (b/a) which is shown in Fig.~\ref{PSF}(c). All of the spots are close to circular. The mean value of $b / a$  is $0.86$. The most circular fluorescent spot has $b/a$ = 0.95, and  the least circular spot $b/a$ = 0.72. 

The size uniformity of the fluorescent spots is reflected in the box-plot analysis of the major and minor axis widths in Fig.~\ref{PSF}(d). The mean value of major and minor axis widths, marked by white lines, are  $\langle a \rangle$ = 6.1 \textmu m and $\langle b \rangle$ = 5.2 \textmu m. The black whiskers show the maximum and minimum of the ensemble. The edges of the blue and red shaded regions denote the upper and lower quartile ($25\%$ and $75\%$) of the data, which are 5.7 \textmu m and 6.4 \textmu m for the major axis $a$, 4.9 \textmu m and 5.4 \textmu m for the minor axis $b$. The green dashed line indicates the diffraction limit, i.e., the width of the PSF in (a). Not only do all of the fluorescent spots have a uniform size, but they are also significantly smaller than the diffraction-limited spot size. Therefore, the spatial resolution enhancement provided by delta speckles is homogeneous.


\begin{figure}[ht]
\begin{center}
\includegraphics[width=3.4in]{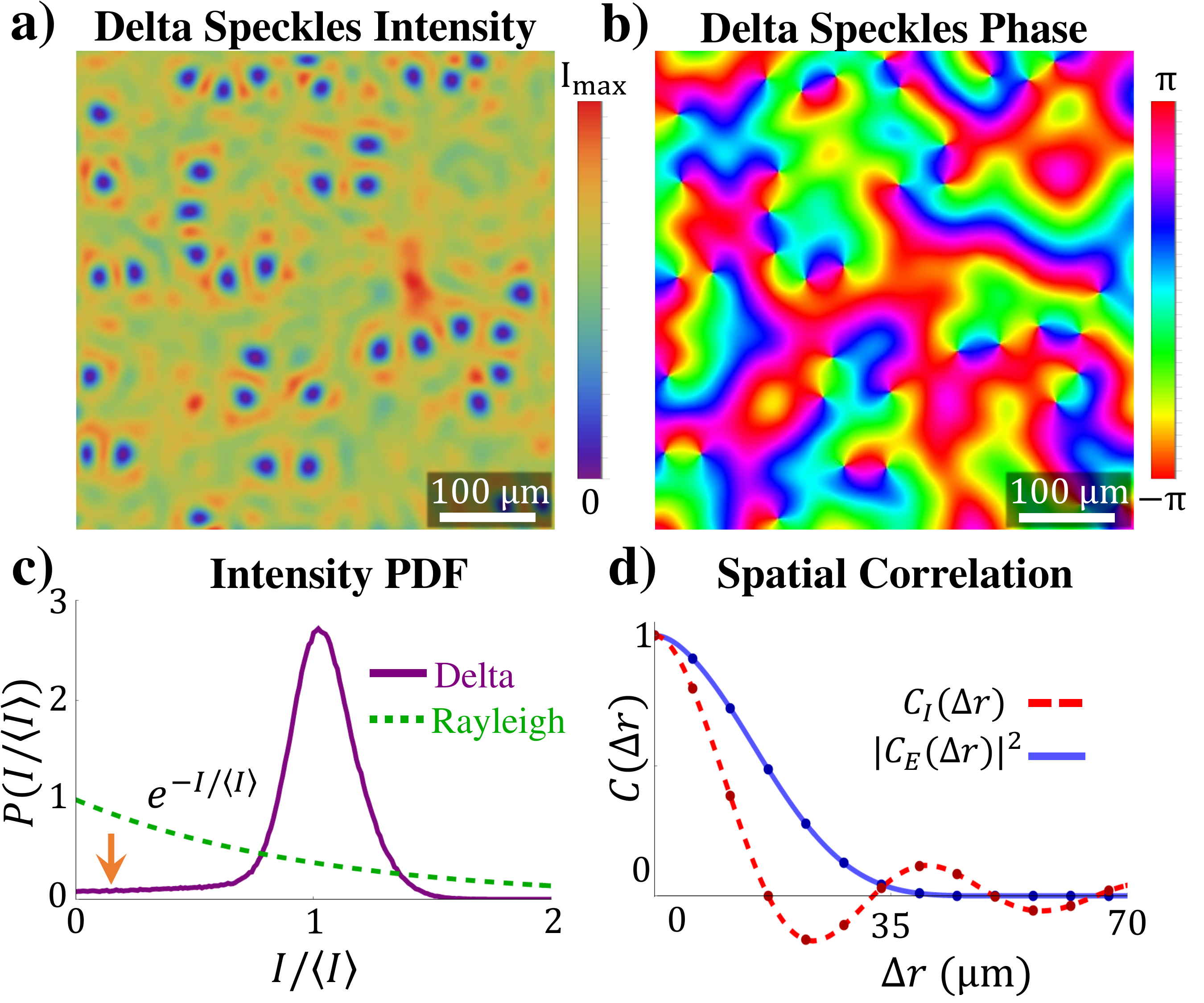}
\caption{\label{SpeckleStat} {\bf Customized Speckle Statistics.} An example delta speckle pattern is presented in (a) and the corresponding phase distribution of its complex field is shown in (b). The phase is randomly distributed between $-\pi$ and $\pi$, indicating that the speckle pattern is fully developed. The intensity PDF (purple solid line) of the delta speckle pattern is shown in (c) and compared with a Rayleigh PDF (green dashed line). The spatial correlation functions of the intensity (red dashed line) and the field (blue solid line) in the delta speckle pattern are shown in (d). The spatial fluctuations of the intensity are faster than those of the field. The anti-correlation ($C_I < 0$) originates from the bright ring surrounding each vortex core. The plots both in (c) and (d) are obtained from an ensemble of 100 speckle patterns.} 
\end{center}
\end{figure}

\section*{Customized Speckle Statistics}

In the previous section, we established the significant advantages of creating isotropic and isomeric vortices in the photoconverting speckle patterns used in fluorescence microscopy.
Here we characterize the customized speckles’ properties in detail and show how their vortex characteristics are transferred to the fluorescent spots. To this end, we use a simpler experimental setup described in the Supplementary Materials. 

An example of an experimentally measured delta speckle pattern is shown in Fig.~\ref{SpeckleStat}(a). The corresponding 2D phase profile of the speckle pattern is shown in (b). While the 2D intensity profile of the speckle pattern is relatively homogeneous apart from the vortices, the corresponding phase profile is random and irregular everywhere. As shown in Supplementary Materials, the complex field of the delta speckles adheres to a circular non-Gaussian PDF. The phase has a uniform probability distribution within the range of $- \pi$ to $\pi$ and the intensity is in dependent of the phase, thus the speckles' field is fully developed.

In Fig.~\ref{SpeckleStat}(c), we compare the intensity PDF of an ensemble of 100 delta speckle patterns like the one shown in (a) (purple line) to a Rayleigh PDF (green dashed line). Because the Rayleigh intensity PDF exhibits an exponential decay, the most probable intensity values in a Rayleigh speckle pattern are close to 0. Therefore, the spatial profile of a Rayleigh speckle pattern is dominated by the low-intensity regions surrounding the optical vortices. The high-intensity regions, which are the bright speckle grains, are sparse, well separated, and isotropic. In many ways the spatial profile of a delta speckle pattern is the inverse of a Rayleigh speckle pattern. Because the spatially uniform regions of high intensity dominate the spatial profile of a delta speckle pattern, the intensity PDF is narrowly peaked and centered around the mean value $\langle I \rangle$. The peak’s width is a reflection of the intensity fluctuations associated with the speckle grains. The presence of optical vortices, and the surrounding dark regions, in the speckle patterns result in a low-intensity tail in the PDF extending to $I=0$ (marked by the orange arrow in (c)). Because the probability density in this tail is low, the vortices are sparse and spatially isolated. 

Using the spatial field correlation function, $C_{E}(\Delta r)$, and the spatial intensity correlation function, $C_{I}(\Delta r)$, (see their definitions in the Supplementary Materials) we can identify the characteristic length scales in a speckle pattern. For a Rayleigh speckle pattern, $C_{I}(\Delta r) = |C_{E}(\Delta r)|^2 $, and therefore both the field and the intensity fluctuate on the same length scale. The spatial correlation length, defined as the full width at half maximum (FWHM) of $C_{I}(\Delta r)$, is determined by the diffraction limit. For a delta speckle pattern, $C_{I}(\Delta r)$ is narrower than $ |C_{E}(\Delta r)|^2 $, as can be seen in Fig.~\ref{SpeckleStat}(d). In the delta speckles, $|C_{E}(\Delta r)|^2 $ remains the same as in Rayleigh speckles, because both patterns are generated by a phase-only SLM and their spatial frequency spectra are identical. Because $C_{I}(\Delta r)$ is narrower than $ |C_{E}(\Delta r)|^2 $ in a delta speckle pattern, it indicates that the intensity varies faster than the field, spatially. Another difference is that with increasing $\Delta r$, the delta speckles’ $C_{I}(\Delta r)$ exhibits a damped oscillation instead of a monotonic decay. These results can be understood as follows: according to the definition of $C_{I}( \Delta r)$, the features in the speckle patterns that contribute the most to the intensity correlation function are those where the intensity deviates the most from the mean value. In a Rayleigh speckle pattern, the bright speckle grains have the greatest difference in intensity relative to the mean value. As a result, the speckle grain size determines the FWHM of $C_{I}( \Delta r)$ in a Rayleigh speckle pattern. Conversely, in a delta speckle pattern, the vortices differ from the mean intensity value more than any other element in the pattern. Therefore, the characteristic features of the optical vortices dictate $C_{I}( \Delta r)$. For example, the high-intensity halo surrounding each vortex core is reflected in the negative correlation $C_{I}( \Delta r) < 0 $ around $\Delta r = 20$ \textmu m, which corresponds to the distance from the vortex center to the high-intensity halo.

\begin{figure}[tb]
\begin{center}
\includegraphics[width=3.4in]{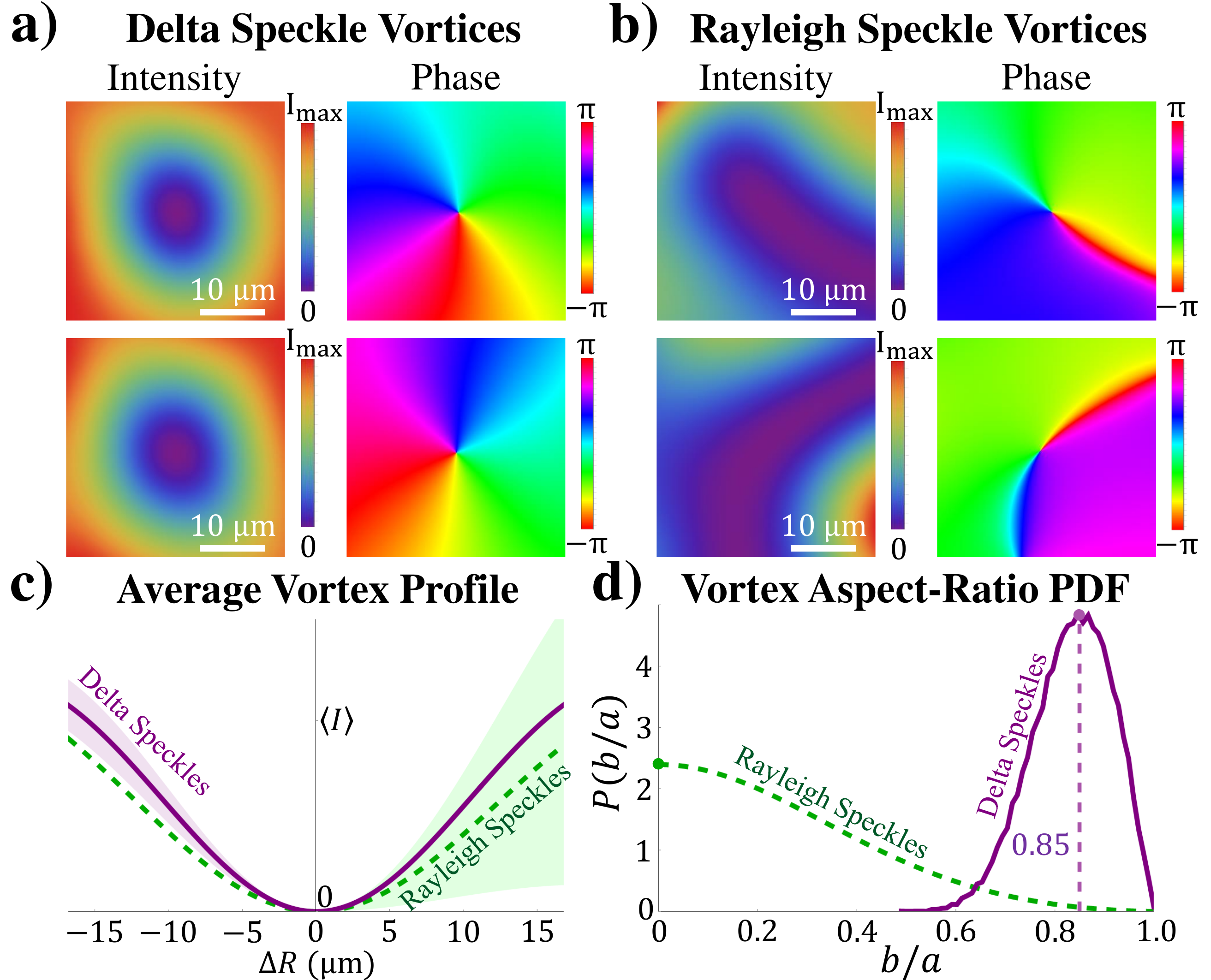}
\caption{\label{Vortex} {\bf Optical Vortex Characteristics.} Two example vortices from a delta speckle pattern, (a), and from a Rayleigh speckle pattern, (b), are shown. While the vortices in the delta speckles are nearly circular, the Rayleigh speckles’ vortices have highly irregular shapes. This property is reflected in (c) where we plot the average intensity profile of light around vortices in 1,000 delta speckle patterns (purple solid line) and around vortices in 1,000 Rayleigh speckle patterns (green dashed line). The edge of the purple and green shaded regions indicates one standard deviation away from the corresponding mean profile. In (d), we plot the probability density of the equal-intensity contours' aspect-ratio around the vortices in 1,000 delta speckle patterns (purple) next to the theoretical prediction for Rayleigh speckle patterns (green).}
\end{center}
\end{figure}

\section*{Vortex Characteristics}
In our photoconversion experiment, effectively, the inverse of the $\lambda = 405$ nm speckle pattern is imprinted onto the uniform fluorescent sample. Therefore after photoconversion, the measured fluorescence originates from the regions of the sample in the vicinity of the optical vortices in the speckle pattern. Here, we investigate the shape and intensity fluctuations of such regions in both delta speckles and Rayleigh speckles. In Fig.~\ref{Vortex}(a,b), we magnify two representative vortex-centered regions in a delta speckle pattern and in a Rayleigh speckle pattern. While the dark region surrounding the vortex cores in the delta speckle pattern is relatively circular, it is elongated and irregular in the Rayleigh speckle pattern. These properties are general and occur within one spatial correlation length from the vortex center. In Fig.~\ref{Vortex}(c), we plot the radial intensity profiles averaged over all vortices in ensembles of delta (purple line) and Rayleigh (green line) speckle patterns. The intensity rises faster with distance from the vortex center in delta speckles, leading to a smaller region of low intensity than found in Rayleigh speckles. In (c) the intensity fluctuations about the average profile are described by the shaded area, whose edges represent one standard deviation from the mean. Relative to Rayleigh speckles, the dramatically reduced intensity fluctuations in the neighborhood of the delta speckles’ vortices leads to more consistent vortex profiles.

The core structure of a vortex is characterized by the equal-intensity contour immediately surrounding the phase singularity.  As with the fluorescent spots analyzed previously, this structure can be described by an ellipse whose major and minor axis lengths are $a$ and $b$. The aspect ratio $b/a$ reflects the degree of isotropy of the vortex core structure. In Fig.~\ref{Vortex}(d), we plot the PDF of $b/a$, $P(b/a)$, obtained from 1,000 independent delta speckle patterns and compare it with that of Rayleigh speckles.  For Rayleigh speckles, $P(b/a)$ has a maximum at $b/a = 0$, indicating the most probable shape of the intensity contour around a vortex is a line \cite{dennis2001, WangPRL2005, GenackPRL2007}. Furthermore, the probability of a circular contour, $b/a=1$, vanishes, confirming the absence of isotropic vortices in a Rayleigh speckle pattern. Contrarily, $P(b/a)$ for the delta speckles features a narrow peak centered at $b/a = 0.85$, meaning all vortices are nearly circular. For comparison, the probability that a vortex will have $b/a > 0.6$ is $99.7\%$ in a delta speckle pattern, while the probability is only $6\%$ in a Rayleigh speckle pattern.  Due to the circular symmetry exhibited by the optical vortices \cite{de2018singular} in delta speckles, the nearest-neighbor vortex spacing, $\cong 45$ \textmu m, is larger than that in Rayleigh speckles, $\cong 27$ \textmu m. Therefore, the average vortex density is lower in a delta speckle pattern, when compared to a Rayleigh speckle pattern.
\begin{figure*}[ht]
\begin{center}
\includegraphics[width=\textwidth]{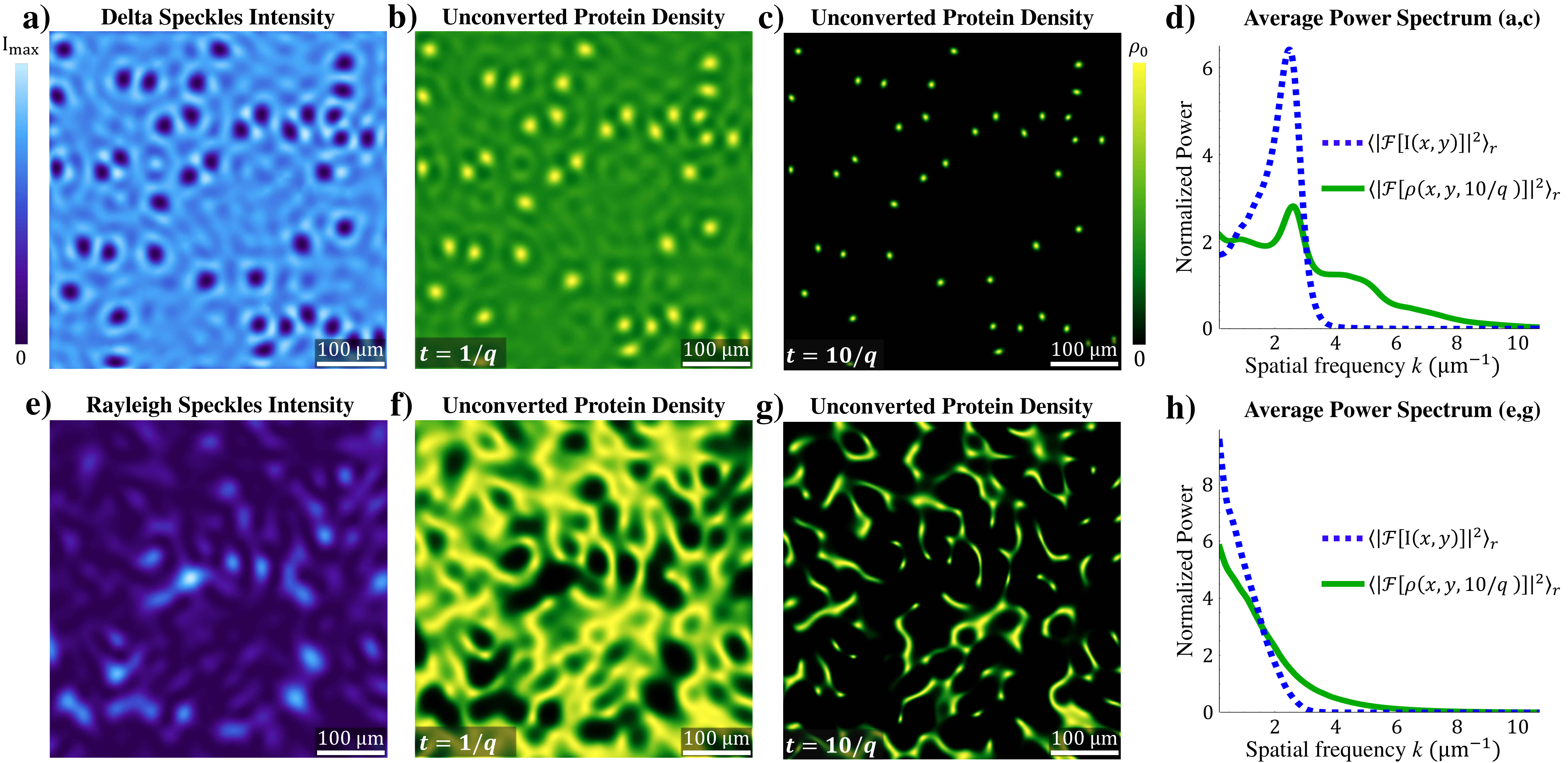}
\caption{\label{Photoconversion} {\bf Photoconversion simulation.} When a delta speckle pattern is used for photoconversion (a), the unconverted regions near the vortices quickly evolve into isotropic and isomeric islands as seen in (b) at $t=1/q$. At long timescales the islands remain, as shown in (c) where $t=10/q$, yet they have considerably smaller size  and higher spatial frequencies (d). When a Rayleigh speckle pattern is used for photoconversion (e), an interconnected web forms instead of isolated islands at $t=1/q$ as shown in (f). Even at long time scales isolated and isotropic islands are rarely seen in (g) and fewer high-spatial frequencies are present when compared to (c), as shown in (h). The power spectra in (d,h) are ensemble-averaged over 100 realizations, and normalized to have a mean value of 1 over the spatial frequencies plotted. Note, the values at $k=0$ are not plotted.}
\end{center}
\end{figure*}
\section*{Protein photoconversion simulation}

In this section, we simulate the photoconversion process with a customized speckle pattern. We consider the case of a uniform sample photoconverted by an intensity pattern $I_p({\bf r})$, starting from $t=0$. The unconverted protein density $\rho({\bf r}, t)$ satisfies the rate equation:
\begin{equation}
\frac{d \rho({\bf r}, t)}{dt} = -q I_p({\bf r}) \rho({\bf r}, t)
\end{equation}
where $q$ is a coefficient describing the photoconversion strength. The solution to this equation is given by
\begin{equation}
\rho({\bf r}, t) = \rho_{0} e^{-q \, I_p({\bf r}) \, t},
\end{equation}
 where $\rho_{0} = \rho({\bf r}, 0)$ is the initial uniform protein density.

The fluorescence intensity from the unconverted proteins $I_e({\bf r}, t)$ is proportional to $\rho({\bf r}, t)$. Thus the negative of the photoconverting intensity  $I_p({\bf r}, t)$ is nonlinearly (exponentially) imprinted onto the fluorescence image $I_e({\bf r}, t)$. In Fig.~\ref{Photoconversion}, we plot $\rho({\bf r}, t)$ for both delta speckles (Fig.~\ref{Photoconversion}(a)) and Rayleigh speckles (Fig.~\ref{Photoconversion}(e)). With increasing photoconversion time the unconverted regions shrink. As can be seen in Fig.~\ref{Photoconversion}(b,c), for delta speckles the unconverted regions evolve to isolated circular spots of homogeneous size. For Rayleigh speckles, the unconverted regions have irregular shapes and highly inhomogeneous sizes (Fig.~\ref{Photoconversion}(f,g)).  We compute the power spectra by Fourier transforming the speckled intensity patterns in (a,e) and the unconverted protein density distributions in (c,g). The azimuthally-averaged power spectra are shown in (d,h). Because delta speckles possess non-local correlations, which shorten the spatial intensity correlation length, the speckles’ intensity power spectrum contains higher spatial frequencies than those present in Rayleigh speckles: which can be seen by comparing the blue dashed lines in (d,h). As a result, the delta speckles produce higher spatial frequencies in the unconverted protein density patterns, especially at long photoconversion time scales, as shown by green solid lines in (d,h).

Our model demonstrates that the circularity of the low-intensity region surrounding each vortex, in a delta speckle pattern, directly translates to the circularity of the corresponding fluorescent spots in the photoconverted sample. This supports our experimental findings, where the aspect ratio of the fluorescent spots has a mean value of $\langle b/a\rangle=0.86$, which agrees with the mean aspect ratio of the optical vortices, $\langle b/a \rangle = 0.85$. Similarly, this model explains the homogeneity in size and shape of the fluorescent spots when a delta speckle pattern is used for photoconversion. When a Rayleigh speckle pattern is used for photoconversion, the vanishing likelihood of having circular vortices coupled with the irregular shape of the surrounding low intensity regions results in the scarcity of isotropic fluorescent spots.
\begin{figure}[ht]
\begin{center}
\includegraphics[width=3.4in]{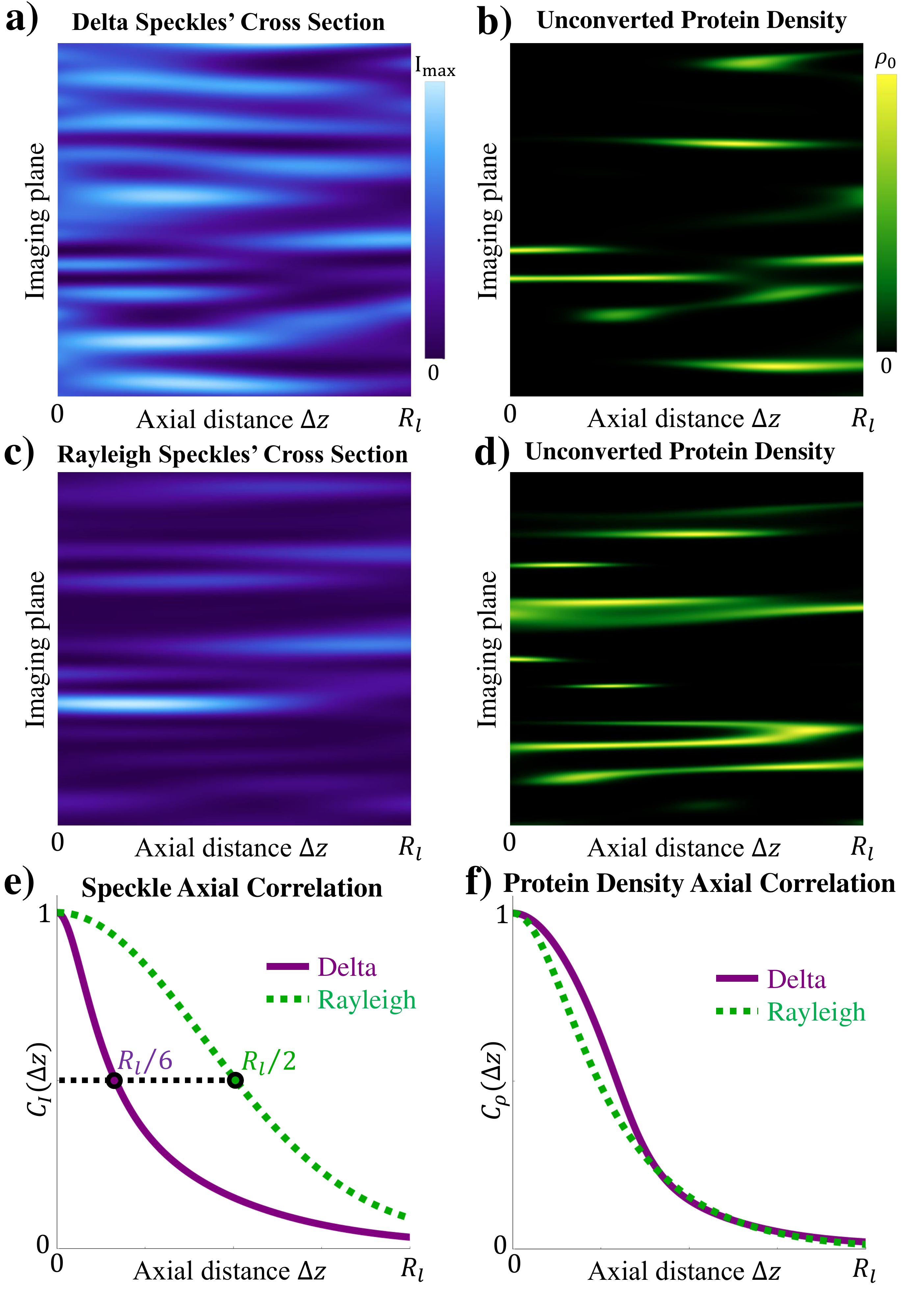}
\caption{\label{PropagationPC} {\bf Axial Propagation.} Axial intensity cross-sections, $I(x=x_{0},y,\Delta z )$, of both delta (a) and Rayleigh (c) speckles are juxtaposed with a simulation of the corresponding unconverted protein density in a uniform sample, $\rho(x=x_{0},y,\Delta z , 10/q)$, (b,d) over one Rayleigh axial-decorrelation length $R_{l}$. In (e), the axial intensity correlation function $C_I(\Delta z)$ of the delta speckles (purple line) is three times narrower than that of the Rayleigh speckles (green dashed line). In (f), the axial correlation function of the unconverted protein density $C_{\rho}(\Delta z)$ generated by delta speckles (purple line) has almost same width as that generated by Rayleigh speckles (green dashed line). We ensemble average over the propagation of 100 speckle patterns to create the curves in (e,f).} 
\end{center}
\end{figure}
\section*{Optical sectioning}

 The rapid axial-decorrelation of a speckled intensity pattern, which enables parallel 3D nonlinear patterned-illumination microscopy, is the result of the field’s uniform phase distribution. The axial decorrelation length of a speckle pattern is defined as the FWHM of the axial intensity correlation function $C_I(\Delta z)$. For a Rayleigh speckle pattern, the axial decorrelation length $R_{l}$ is proportional to the Rayleigh range: which is determined by the axial diffraction limit $2 \lambda/ {\rm NA}^{2}$. With axial propagation, a delta speckle pattern evolves into Rayleigh speckles over one $R_{l}$. In Fig.~\ref{PropagationPC} (a), an example axial intensity cross-section, $I(x=x_{0},y,\Delta z )$, of a delta speckle pattern over one $R_{l}$ is juxtaposed with a numerical simulation of the corresponding unconverted protein density in a uniform sample, $\rho(x=x_{0},y,\Delta z , t = 10/q)$. For comparison, the simulation results for a Rayleigh speckle pattern are shown in (c,d). As the illuminating speckle patterns decorrelate axially, the fluorescent spots (corresponding to optical vortices) move laterally, enabling parallel 3D nonlinear patterned-illumination microscopy.

 As shown in Fig.~\ref{PropagationPC} (e), the axial intensity correlation function $C_I(\Delta z)$ of the delta speckles has a notably narrower width than that of the Rayleigh speckles. Consequently, the axial decorrelation length of the delta speckles (FWHM of the axial intensity correlation function) is $R_{l}/3$, which is three times shorter than that of the Rayleigh speckles. This behavior can be attributed to the rapid axial evolution of a delta speckle pattern into a Rayleigh speckle pattern, a general feature of  customized speckles \cite{benderAPL2019, benderOE2019, bender2018} (see the Supplementary Materials for details). However, the axial decorrelation lengths of the unconverted protein density in uniform samples, $\rho({\bf r}, t)$, remains nearly identical for long exposure times  (Fig.~\ref{PropagationPC} (f)). This is because the axial decorrelation of $\rho({\bf r}, t)$ is dictated by the lateral movement of optical vortices, when propagating axially, which is almost the same for both families of speckles. Since optical vortices in a delta speckle pattern are farther apart than those in a Rayleigh speckle pattern, their transverse motion is slightly slower, leading to a minor broadening of $C_{\rho}(\Delta z)$. Nevertheless, the delta speckle illumination provides approximately the same optical sectioning as the Rayleigh speckle.

\section*{Discussion and Conclusion}

In summary, we have presented a proof-of-principle demonstration of parallel pattern illumination with a family of tailored speckle patterns. By customizing the statistical properties of speckle patterns for photoconversion, we obtain a spatial resolution that is three times higher than the diffraction limit of the illumination optics. The isometric and circular vortices in the tailored speckle patterns provide a homogeneous and isotropic spatial resolution enhancement: which cannot be obtained from standard Rayleigh speckles.

Since the photoconversion process is nonlinear, in principle, there is no limit on the spatial resolution that can be reached. In reality, because the intensity of our photoconverting laser beam is relatively low, the photoconversion process takes a long time ($\sim 12$ hours). During this time, sample drift and protein motion limit the spatial resolution that can be achieved with fluorescence from the unconverted regions. A further increase in spatial resolution is possible using a higher-powered laser with dilute samples of immobile proteins.

In the photoconversion experiment, the illumination optics has a relatively low NA so that the fluorescence spots of size exceeding 4 \textmu m can be well resolved with the detection optics of 1.1 \textmu m resolution. To reach nanoscale resolution, high-NA optics is required to create the photoconverting speckle pattern. In this case, the vector nature of the light field must be considered. It is known that the axial field at a vortex center can be canceled by manipulating the polarization state of light \cite{hao2010phase, 2012_Galiani_OE}. In a Rayleigh speckle pattern, the intensity contours around a vortex core are elliptical, and the polarization state must be an ellipse with an identical aspect ratio and an identical handedness in order to cancel the axial field \cite{2016_Guillon_PRL, 2019_Guillon_NatCommun}. Since the elliptical contours vary in aspect ratio from one vortex to the next in a Rayleigh speckle pattern, it would be challenging, if not impossible, to set the polarization state of a speckle pattern to cancel the axial fields at all of its vortices. In a delta speckle pattern, in contrast, the vortices are almost circular, and half of them have the same handedness. With circularly polarized light, the axial fields  will vanish at half of the vortices: specifically those with the same handedness. To be consistent with high-NA applications, delta speckles with circular polarization are generated in our experiment.

In our experiment, a uniform film of purified protein is photoconverted by a speckle pattern. In the Supplementary Materials we show that this technique works with live yeast cells, which are nonuniformly distributed in space. 

Finally, 2D raster scanning delta speckles over a fluorescent sample enables 3D superresolution imaging of the sample: via the same technique demonstrated with Rayleigh speckles in \cite{2019_Guillon_NatCommun}.  Alternatively, it is possible to eliminate scanning from this process by illuminating the entire sample with multiple, distinct, delta speckle patterns which cover the field of view. While the axial resolution obtained using delta speckles for illumination is comparable to that of Rayleigh speckles, further improvements can be made. Specifically, accelerating the transverse motion of optical vortices in delta speckles can enhance their optical sectioning capabilities.

\section*{Acknowledgments}
We thank Yaron Bromberg for insightful comments. We thank Meadowlark Optics (https://www.meadowlark.com/) for providing the spatial light modulator. 

\section*{Declarations}
J.B. has financial interests in Bruker Corp. and Hamamatsu Photonics. The rest co-authors declare no competing interests.

\section*{Funding}
This work is supported partly by the Office of Naval Research (ONR) under Grant No. N00014-20-1-2197, and by the National Science Foundation under Grant No. DMR-1905465, the Wellcome Trust (203285/B/16/Z) and the National Institutes of Health (NIH P30 DK045735). MS was supported by the National Institute of General Medical Sciences of the National Institutes of Health under award number R01GM026132 and R01GM118486, and the Wellcome Trust (203285/B/16/Z)


\bigskip \noindent See \href{link}{Supplement 1} for supporting content.
	
\bibliography{Bibliography}
\end{document}


\title{Circumventing the optical diffraction limit with customized speckles: supplementary material}
\author{Nicholas Bender,\authormark{1} Mengyuan Sun,\authormark{2} Hasan Y{\i}lmaz,\authormark{1} Joerg Bewersdorf,\authormark{3,4} and Hui Cao\authormark{1,*}}
\address{\authormark{1} Department of Applied Physics, Yale University, New Haven CT 06520, USA\\
\authormark{2} Department of Molecular Biophysics and Biochemistry, Yale University, New Haven CT 06520, USA \\
\authormark{3} Department of Cell Biology, Yale University School of Medicine, New Haven CT 06520, USA\\
\authormark{4} Department of Biomedical Engineering, Yale University, Yale University, New Haven CT 06520, USA
}

\email{\authormark{*} hui.cao@yale.edu} 

\begin{abstract}
	This document provides supplementary information to “Circumventing the optical diffraction limit with customized speckles.” In the first section, we outline our method of preparing the uniform mEos gel-samples. In the second section we discuss our method for analyzing the unconverted fluorescent spots present in the photoconverted samples. In the third section, we outline the method we use to create delta speckles. In the fourth section, we conduct a photoconversion fluorescence microscopy demonstration with live yeast cells. In the fifth section, we describe the second experimental setup we used to characterize the statistical properties of the delta speckles. In the sixth section we provide the definitions we use to calculate the spatial field and intensity correlations. In the seventh section we describe our method for analyzing the optical vortexes in the speckle patterns. In the eighth section we present the complex-field PDF of the delta speckles. In the nineth section we discuss the delta speckles’ axial evolution.
\end{abstract}
	\section{Purified Protein Sample Preparation}  
	The plasmid for mEos3.2 expression is cloned with PCR and NEBuilder assembly, and transformed into BL21-CondonPlus (DE3) competent cells. The mEos3.2 protein is purified using Ni-NTA His-Bind resin and dialyzed in dialysis buffer (20 mM Tris, pH 7.5/ 10 mM NaCl/ 1mM EDTA/ 10 mM BME). To immobilize mEos3.2, a mixture containing 42 \textmu L purified mEos3.2 (0.1 mM), 30 \textmu L 30.8 $\%$ Acrylamide/bis-acrylamide, 0.5 \textmu L 10 $\%$ APS, and 0.5 \textmu L TEMED is sandwiched between a clean coverslip and a slide to make a 25-\textmu m-thick gel.

	\section{Unconverted Fluorescence Analysis}
	
	We use the following procedure to quantitatively analyze the size and shape of the fluorescent spots created by the optical vortices in a photoconverting delta speckle pattern. First, we locate the center of each fluorescent spot: ${\bf r}_c = \int {\bf r} \, I(\bf r) d{\bf r} / \int I(\bf r) d{\bf r}$. Using the 38 \textmu m $\times$ 38 \textmu m region surrounding each fluorescent spot, which corresponds to 61 $\times$ 61 pixels on the CCD camera, we numerically construct a 2D interpolation of every fluorescent spot. Next, we rotate each interpolated grid around the center  ${\bf r}_c$ in increments of $1^{\circ}$, over a total of $180^{\circ}$, and record the 1D profile along the horizontal axis for each rotation. Then we fit each of these 1D profiles to a Gaussian function, 
	$I_a \times \exp[-(x-x_0)^{2}/(2 \sigma^{2})] + I_c$, 
	with $I_a$, $x_0$, $\sigma$, $I_c$ as the fitting parameters. After determining the maximum and minimum values of $\sigma$ from all rotation angles, for each fluorescent spot, we extract both the major and minor axis widths from the full width at half maximum (FWHM)  $2 \sigma \sqrt{2 \ln(2)}$. For the results shown in Fig. 2 of the main text, the average $R^{2}$ fitting coefficient is $0.996$.  Generally the rotation angles corresponding to the maximum and the minimum widths are offset by $90^{\circ}$, reflecting the elliptical shape of the fluorescent spots.
	
	\section{Delta Speckle Generation Method}
	
	A delta speckle pattern can be generated following the speckle customization method outlined in our previous paper \cite{benderAPL2019}. In this process, first, a Rayleigh speckle pattern is numerically generated by applying a random phase pattern with a uniform probability density $(-\pi, \pi]$ to the experimentally measured field-transmission matrix of our wavefront shaping system. The recorded Rayleigh speckle pattern’s intensity at every spatial location is numerically replaced by a constant value (equal to the mean intensity), while the phase is unchanged. This modified pattern has a delta-function intensity PDF and a circular joint PDF for the complex field. Using this pattern as the starting point in a transmission-matrix-based local nonlinear optimization algorithm, specifically the Broyden-Fletcher-Goldfarb-Shanno (BFGS) algorithm \cite{nLopt2}, we search for a phase-only modulation pattern to display on the SLM which generates the target speckle pattern on the camera plane \cite{benderAPL2019, benderOE2019, bender2018}. A perfect match is impossible, however, due to the existence of phase singularities. Instead, local minima are found that correspond to speckle patterns with nearly uniform intensity apart from the optical vortices. The resulting intensity PDF has a narrow peak and a tail extending to zero, as denoted by the orange arrow in Fig. 3(c) of the main text. To generate different speckle patterns with this approach, we simply initialize the SLM with different random phase patterns.
	
	\section{Live sample demonstration}
	
	In the main text, we use delta speckle patterns to photoconvert a homogeneous film of purified protein to demonstrate the isometric and isotropic spatial resolution enhancement they provide. Here we show that our technique is compatible with living samples, which are typically inhomogeneous.
	
	\subsection*{Live yeast cell culture and preparation.} 
	The \textit{S. Pombe} strain \textit{Leu1::Leu1+ pAct1 mEos3.2 nmt1Term ade6-M216 his3-$\Delta$1 leu1-32 ura4-$\Delta$18} is generated from NruI digested plasmid pJK148-pAct1-mEos3.2-nmt1Term through homologous recombination. Cytosolic mEos3.2 is expressed from the act1 promoter in the endogenous Leu1 locus. Cells are grown in exponential phase at 25 °C in YE5S-rich liquid medium in 50 mL flasks in the dark before switching to EMM5S-synthetic medium for 12-18 hours, to reduce the cellular autofluorescence background. Live cells are concentrated 10- to 20-fold by centrifugation at 3,000 rpm for 30 s and resuspended in EMM5S for imaging. Concentrated cells in 10 uL are mounted on a thin layer consisting of 35 \textmu L 25 $\%$ gelatin (Sigma-Aldrich; G-2500) in EMM5S.
	
	\begin{figure*}[ht]
		\begin{center}
			\includegraphics[width=\textwidth]{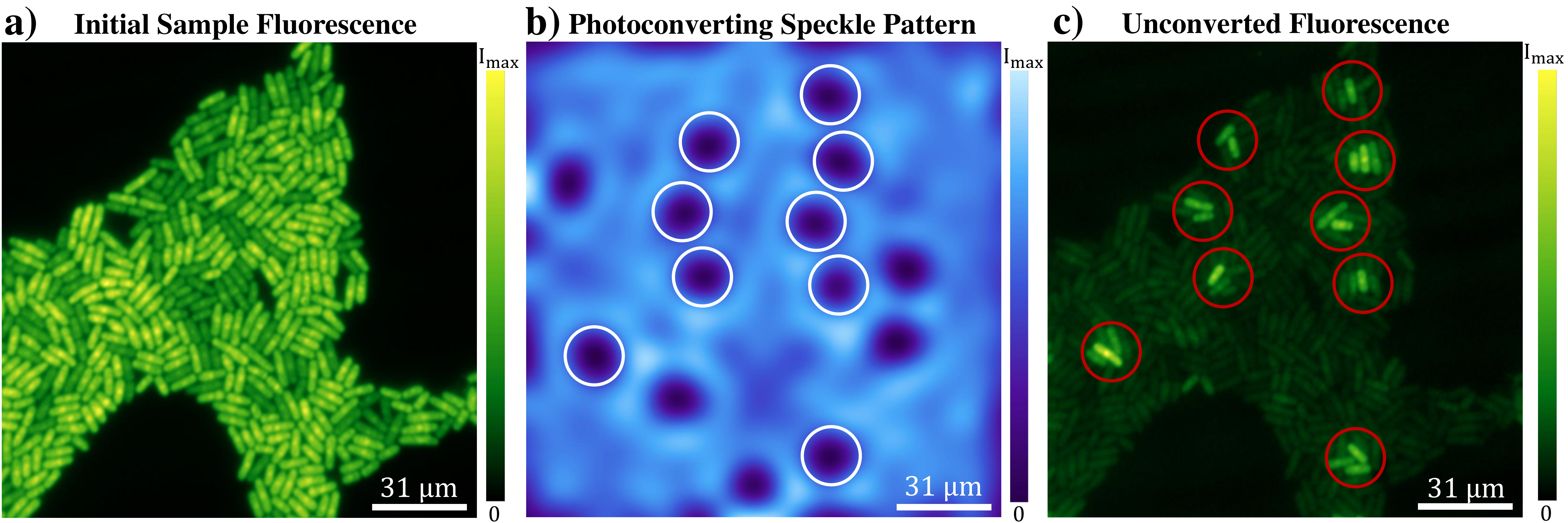}
			\caption{\label{LiveSample} {\bf Photoconversion of live yeast cells with a customized speckle pattern.} In (a), we present an optical image of the fluorescent light emitted from a sample of live yeast cells before they are photoconverted by the delta speckle pattern shown in (b). The white circles in (b) indicate the vortices which overlap with the yeast cells. An image of the fluorescent light emitted by the live-cell sample after photoconversion is shown in (c). The red circles in (c) correspond to the white circles in (b).} 
		\end{center}
	\end{figure*}
	
	\subsection*{Photonconverting live yeast cells with customized speckles.} 
	We illuminate the collection of nonuniformly distributed yeast cells shown in Fig.~\ref{LiveSample} (a) with the photoconverting speckle pattern presented in (b). The optical vortices of the delta speckle pattern do not photoconvert the yeast cells in their vicinity. Consequently after photoconversion, multiple isolated groups of cells will emit fluorescence when illuminated by 488 nm light. In the fluorescence image taken after photoconversion, the fluorescent regions (marked by red circles in (c)) have a one-to-one correspondence to the optical vortices -in the delta speckle pattern- that overlap with the yeast cells (marked by white circles in (b)). Because the mEos3.2 protein is cytosolic, it is impossible to obtain fluorescence from a sub-cellular region, even if we shrink the dark region surrounding each vortex core using high-NA optics. Nevertheless, we are able to select individual groupings of live cells located near the vortices.

	\section{Speckle Characterization Setup}
	
	We use the following setup to characterize the statistical properties of the delta speckles (Figs. 3-5). A linearly-polarized CW laser beam with a wavelength of $\lambda = 642$ nm illuminates a phase-only reflective SLM (Hamamatsu LCoS X10468). The pixels on the SLM modulate the incident light’s phase between $0$ and $2 \pi$: in increments of $2 \pi / 170$. We bypass the unmodulated light by writing a binary diffraction grating with a period of 8 pixels on the SLM and work with light in the first diffraction order. Macropixels are formed by groups of $16 \times 16$ pixels, each containing the binary diffraction grating. To generate a speckle pattern in the far-field, we modulate the square array of $32 \times 32$ macropixels in the central part of the phase modulating region of the SLM. Outside the central square, we display an orthogonal phase grating on the remaining illuminated pixels to diffract the laser beam away from the CCD camera (Allied Vision Prosilica GC660). The SLM and the CCD camera are placed on opposite focal planes of a lens ($f=500$ mm). In this setup, the full width at half-maximum of a diffraction-limited focal spot is 29 \textmu m.

	\section{Definition of the spatial field/intensity correlation function}
	The spatial correlation function of the speckles' field is defined as:
	\begin{equation} 
	C_{E}(\Delta {\bf r}) \equiv \frac{\langle E({\bf r}) E^{*}({\bf r} + \Delta {\bf r})\rangle}{\sqrt{\langle |E({\bf r})|^2\rangle}\sqrt{\langle |E({\bf r} + \Delta {\bf r})|^2\rangle}}
	\end{equation}
	where $\langle ... \rangle$ denotes spatial averaging over $\bf r$. 
	
	The spatial correlation function of the speckles' intensity is defined as:
	\begin{equation} 
	C_{I}(\Delta {\bf r})  \equiv  \frac{\langle \delta I({\bf r}) \delta I({\bf r} + \Delta {\bf r})\rangle}{\sqrt{\langle [\delta I({\bf r})]^2 \rangle}\sqrt{\langle [\delta I({\bf r} + \Delta {\bf r})]^2\rangle}} \\
	\end{equation}
	where $\delta I({\bf r}) = I({\bf r})-\langle I({\bf r})\rangle$ denotes intensity fluctuations around the mean. Because the spatial correlation functions of the speckle patterns studied in this work are azimuthally symmetric and depend only on the distance $\Delta r = |\Delta {\bf r}|$; $|C_{E}(\Delta {\bf r})|^2$ and $C_{I}(\Delta {\bf r})$ can be represented by $|C_{E}(\Delta r)|^2$ and $C_{I}(\Delta r)$.

	\section{Techniques For Characterizing Optical Vortices}
	We use the method outlined in Ref.~\cite{Angelis2016, de2018singular} to locate the optical vortices in our speckle patterns. We use the method outlined in Ref.~\cite{GenackPRL2007} to fit the major and the minor axes of the constant intensity contours and compute the aspect ratios. The intensity profiles in Fig. 4(c) of the main text are obtained by taking the 1D intensity cross section along the horizontal axis of every vortex in 1,000 Rayleigh speckle patterns and 1,000 delta speckle patterns.
	
	\section{The joint PDF of the complex speckle fields}
	
	\begin{figure}[hthb]
		\begin{center}
			\includegraphics[width=3.4in]{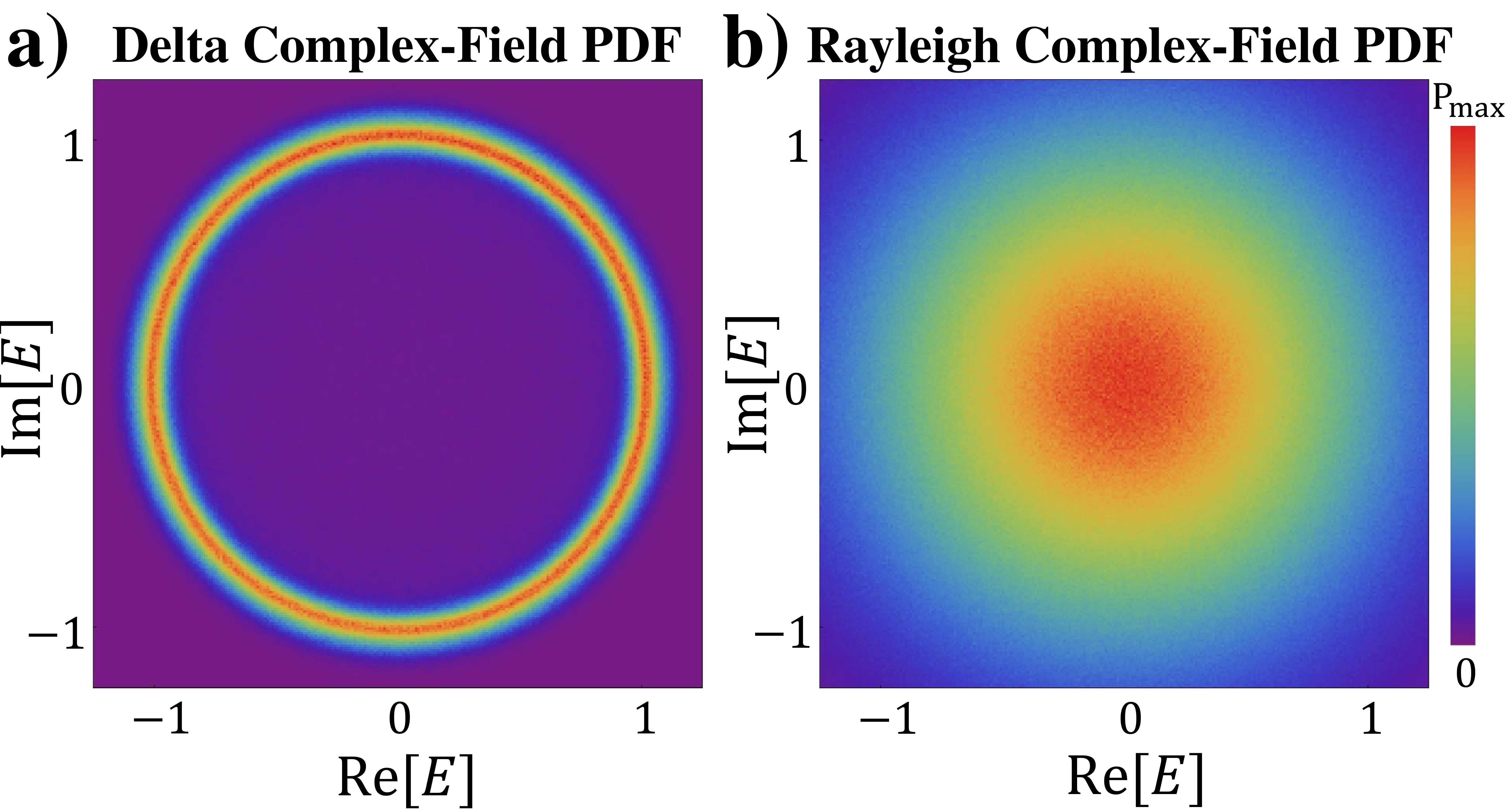}
			\caption{\label{FieldPDF} {\bf Joint PDF of the complex speckle fields.} In (a), we show the circular non-Gaussian joint PDF associated with delta speckles next to the circular Gaussian joint PDF of Rayleigh speckles in (b). In both cases, the speckles are fully developed.}
		\end{center}
	\end{figure}
	
	A speckle pattern is ‘fully developed’ if the joint PDF of its complex field is azimuthally invariant. This means that the phase distribution of the speckle pattern is uniform between $0$ and $2 \pi$, and that the amplitude values of the field are independent of the phase values. This means that the amplitude and phase distributions of a fully-developed speckle pattern are statistically independent. In Fig.~\ref{FieldPDF}, we plot the complex field PDF calculated from 1,000 Rayleigh speckle patterns and 1,000 delta speckle patterns. While the field PDF obeyed by Rayleigh speckles is a circular Gaussian function, the delta speckles adhere to a circular non-Gaussian function. Because the field PDF is circular for delta speckles, they are fully developed like Rayleigh speckles.

	\section{Speckle decorrelation upon axial propagation}
	\begin{figure}[ht]
		\begin{center}
			\includegraphics[width=3.4in]{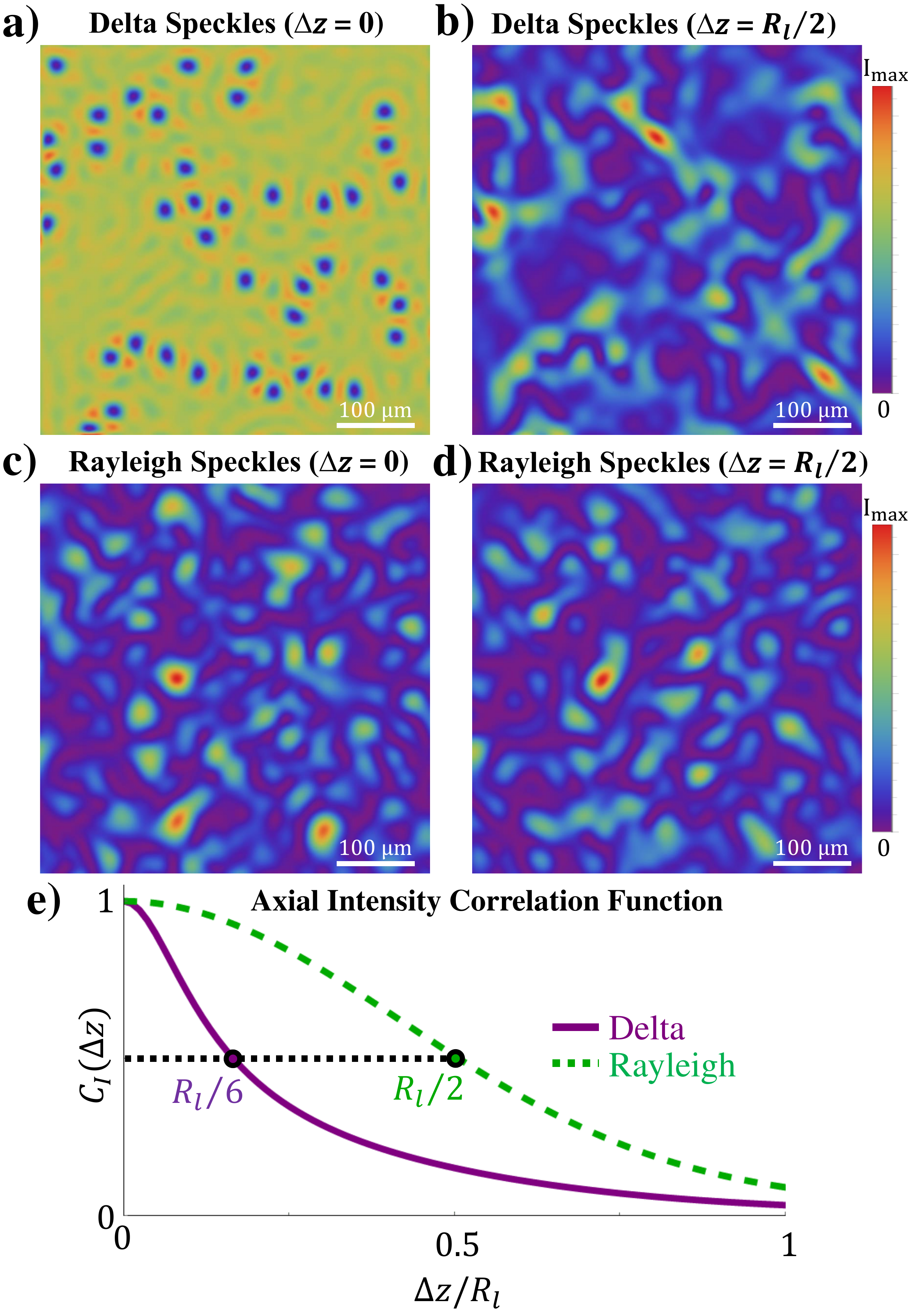}
			\caption{\label{Prop} {\bf Axial decorrelation of speckle patterns.} An example delta speckle pattern at $\Delta z=0$ and $\Delta z=R_{l}/2$ is shown in (a) and (b), while an example Rayleigh speckle pattern at $\Delta z=0$ and $\Delta z=R_{l}/2$ is shown in (c) and (d). In (e), the axial intensity correlation function $C_I(\Delta z)$ for delta speckles (purple line) and Rayleigh speckles (green dashed line) is shown. We ensemble average over the propagation of 100 speckle patterns to create the curves in (e).}
		\end{center}
	\end{figure}
	The rapid axial decorrelation of a fully-developed speckle pattern enables parallel 3D nonlinear patterned-illumination microscopy. The axial intensity correlation function is defined as 
	\begin{equation} 
	C_{I}(\Delta z)  \equiv  \frac{\langle \delta I({\bf r}, z_0) \delta I({\bf r} , z_0 + \Delta z)\rangle} {\sqrt{\langle [\delta I({\bf r}, z_0)]^2 \rangle}\sqrt{\langle [\delta I({\bf r}, z_0 + \Delta z)]^2\rangle}} \\
	\end{equation}
	where $z_0$ is the axial position of the focal plane, $\langle ... \rangle$ denotes averaging over transverse position $\bf r$, and $\delta I({\bf r}, z) = I({\bf r}, z)-\langle I({\bf r}, z)\rangle$ is the intensity fluctuation around the mean.
	The FWHM of $C_I(\Delta z)$ defines the axial decorrelation length. For a Rayleigh speckle pattern, the axial decorrelation length $R_{l}$ is proportional to the Rayleigh range: which is determined by the axial diffraction limit $2 \lambda/ {\rm NA}^{2}$.
	
	\begin{figure*}[t]
		\begin{center}
			\includegraphics[width=\textwidth]{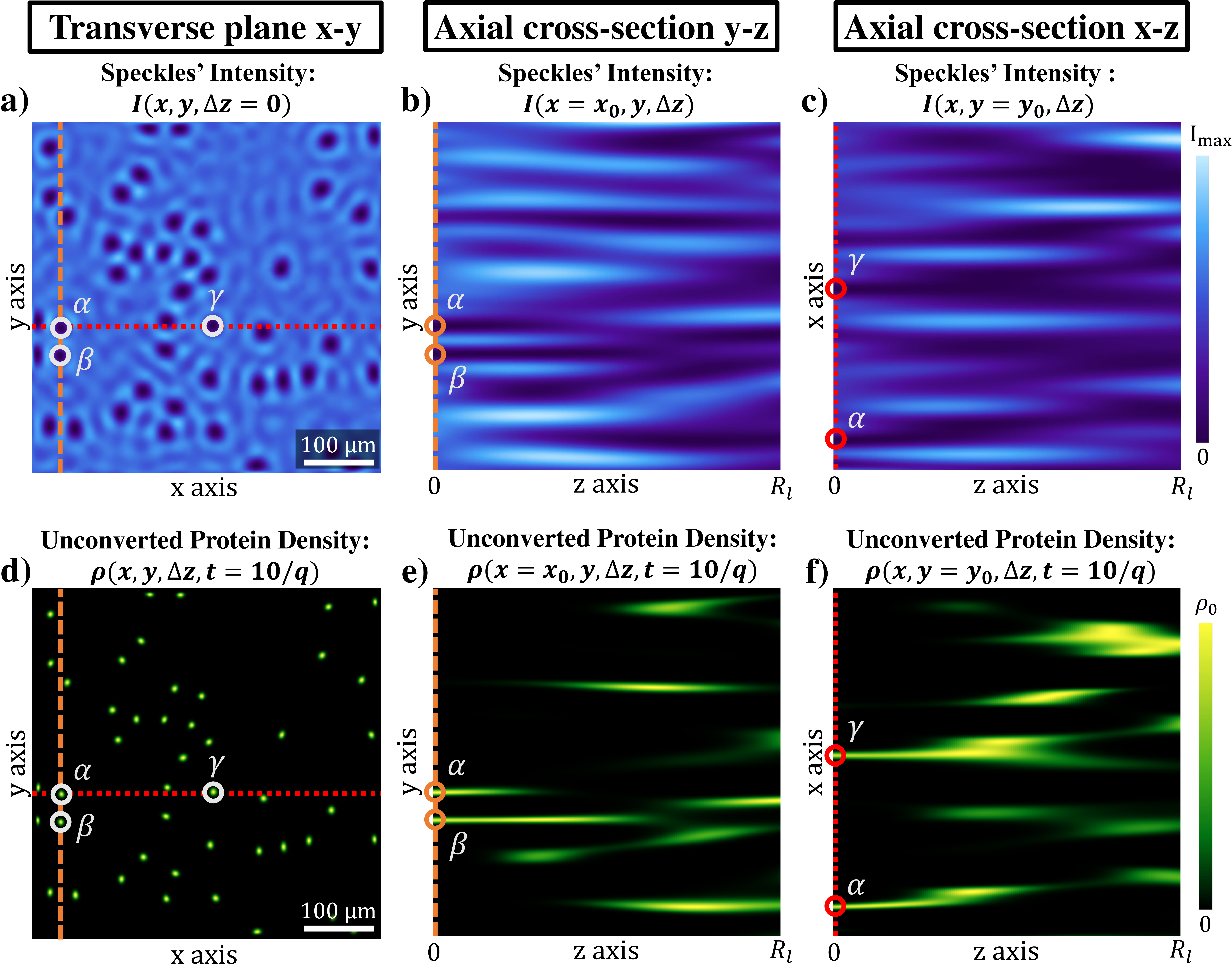}
			\caption{\label{Spot} {\bf Three-dimensional photoconversion with delta speckles.} The calculated spatial distribution of a delta speckle pattern's intensity on a transverse plane at $z= 0$, $I(x,y,\Delta z = 0)$, is shown in (a) along with the intensity distribution over two axial cross-sections, $I(x=x_{0},y,\Delta z )$ in (b) and $I(x,y=y_{0},\Delta z )$ in (c). A photoconversion simulation using the speckles in (a-c) gives the corresponding distribution of the unconverted protein density in a uniform sample, $\rho({\bf r}, t)$, in (d-f).}
		\end{center}
	\end{figure*}
	
	We numerically simulate the axial propagation of delta speckles and compare it with that of Rayleigh speckles. In Fig.~\ref{Prop}, an example delta speckle pattern at $\Delta z=0$ and $\Delta z=R_{l}/2$ is shown in (a) and (b), while an example Rayleigh speckle pattern at $\Delta z=0$ and $\Delta z=R_{l}/2$ is presented in (c) and (d). A qualitative comparison between (a), (b) and (c), (d) illustrates that the delta speckles’ transverse intensity-profile axially evolves much faster than the Rayleigh speckles’ profile. We plot the axial correlation function of the delta speckles (purple solid line) and the Rayleigh speckles in (e). The axial decorrelation length of the delta speckles (FWHM of axial intensity correlation function) is $R_{l}/3$, which is three times shorter than that of the Rayleigh speckles.
	
	In Fig. \ref{Spot} we show the effect the transverse motion of optical vortices in delta speckles -upon axial propagation over one $R_{l}$- has on the fluorescent spots in a photoconverted sample. An example delta speckle pattern, $I(x,y,\Delta z = 0)$, is shown in (a) along with two vertical cross-sections, $I(x=x_{0},y,\Delta z )$ in (b) and $I(x,y=y_{0},\Delta z )$ in (c). The axial cross-sections show the speckle’s intensity, along a line in $x$ or $y$ axis, as a function of axial propagation from $\Delta z=0$ to $\Delta z=R_{l}$.  Using Eq. 2 from the main text, we simulate the corresponding unconverted protein density, $\rho({\bf r}, t)$, in a sample after it is photoconverted with the speckle pattern shown in (a-c). In (d-f) we show the unconverted protein densities after the sample is photoconverted for $t=10/q$. Not only do the fluorescent spots located at ${\bf r} = \alpha, \beta, \gamma$ move transversely and leave the axial cross-section in (e) \& (f) before propagating to $\Delta z=R_{l}$, but also, other fluorescent spots enter and exit the axial cross-sections: demonstrating that delta speckles can be used to obtain 3D super-resolution images.

	\bibliography{Bibliography}